# Error-Free and Current-Driven Synthetic Antiferromagnetic Domain Wall Memory Enabled by Channel Meandering

Pengxiang Zhang[1*], Wilfried Haensch[1], Charudatta M. Phatak[1,2], Supratik Guha[1,3]

[1]Materials Science Division, Argonne National Laboratory, Lemont, IL 60439, USA

[2]Department of Materials Science and Engineering, Northwestern University, Evanston, IL 60208, USA

[3]Pritzker School of Molecular Engineering, University of Chicago, Chicago, IL 60637, USA

[*]Corresponding Author: zhangp@anl.gov

**Abstract:** We propose a new type of multi-bit and energy-efficient magnetic memory based on current-driven, field-free, and highly controlled domain wall motion. A meandering domain wall channel with precisely interspersed pinning regions provides the multi-bit capability of a magnetic tunnel junction. The magnetic free layer of the memory device has perpendicular magnetic anisotropy and interfacial Dzyaloshinskii-Moriya interaction, so that spin-orbit torques induce efficient domain wall motion. Using micromagnetic simulations, we find two pinning mechanisms that lead to different cell designs: two-way switching and four-way switching. The memory cell design choices and the physics behind these pinning mechanisms are discussed in detail. Furthermore, we show that switching reliability and speed may be significantly improved by replacing the ferromagnetic free layer with a synthetic antiferromagnetic layer. Switching behavior and material choices will be discussed for the two implementations.

## 1   Introduction

Magnetic non-volatile memory devices are promising for low-power, high-speed storage, and computing[1]. Comparing to the commercialized spin-transfer torque magnetic random-access memory (STT-MRAM), spin-orbit torque magnetic random-access memory (SOT-MRAM) is considered one of the most promising next-generation magnetic memory solutions[2–4]. In SOT-MRAM read and write current path are separated, so that the device's lifespan can be increased by many orders of magnitudes, and due to different magnetic dynamics, the device can be switched much faster than STT-MRAM. However, the SOT-MRAM bit-cell requires two transistors per cell and therefore comes with a density hit compared to the STT-MRAM[5–8]. For the magnetic component of SOT-MRAM, magnetic materials with perpendicular magnetic anisotropy (PMA) allow much higher storage density compared to the ones with in-plane magnetic anisotropy, but the former needs an in-plane static magnetic field to operate, while the latter does not. While many workarounds have been proposed to eliminate this need for SOT-MRAM with PMA, none of them is straightforward and most of them do not scale well[7,8].

Conventional MRAM is a one-bit memory[3]. While NAND Flash memory devices offer a multi-bit storage solution, they are not well suited for off-chip memory (typically DRAM) or cache (typically SRAM) applications due to their limited cyclability and long write time. Memristor type memory devices gained attention as possible solution for multi-bit memory[9]. However, their sensitivity to process variations and limited write endurance, making large scale applications challenging[9,10]. Recently, they have been used in applications with less stringent requirements on precision, like cross-point arrays to perform analog matrix-vector multiplications that are important for deep learning algorithms with the hope of improved computational efficiency[9,10]. On the other hand, MRAM has also been introduced to cross-point arrays to perform analog matrix-vector multiplications, but the MRAM used remains binary[11]. A multi-bit magnetic memory, which operates solely on electric currents and does not require external magnetic fields, is highly valuable for not only non-volatile memories, but also neuromorphic computing.

The key to a multi-bit magnetic memory is the precise control of domain walls, which are thin boundaries separating different magnetic domains. This idea was originated from the concept of racetrack memory[12]. The racetrack memory structure consists of thin magnetic nanowire channels. Bits of data (1s and 0s) are stored in the form of magnetic domains, which are separated by domain walls. For data movement, electric current shifts these domain walls along the "racetrack". Specialized sensors (magnetic tunnel junctions) detect the position of the domain walls (and thus the stored data) as they move past a fixed point on the nanowire via electrical resistance (tunneling magnetoresistance). Writing data involves manipulating the magnetic orientation of domains using current pulses. Now, instead of having a racetrack memory structure, if only one domain wall exists and separates the magnetic channel into two regions of opposite magnetic orientations, and the domain wall moves between two opposite ends continuously, to vary the fraction of the "up" and "down" magnetic domains within the channel, this creates multiple resistance states.

An effective way to drive such domain motion is via the spin-orbit torques (SOT) from an underlying spin current source, which is a non-magnetic conductor converting electrical current to spin current via the Dzyaloshinskii-Moriya Interaction (DMI)[7] interaction in a ultrathin magnetic film with perpendicular magnetic anisotropy (PMA). The spin current flowing through this ultrathin layer can move a domain wall parallel or antiparallel to the direction of the current flow, and this changes the fraction of up and down domains. A separate read mechanism through a magnetic tunnel junction enables reading out the tunneling current through the magnetic layer. Since the tunneling current is proportional to the relative ratios of the two domains, this allows for a multi-level magnetic memory cell.

The control of the domain wall motion is influenced by many factors: material defects, thermal fluctuations, spatial variability, and drive current control are the most important ones[13,14]. Variations in drive current pulses will directly translate to an uncertainty in the domain wall position. The solutions proposed to mitigate this uncertainty are based on introducing a periodic energy pattern that will pin the domain wall at predetermined positions. This can be done by geometrical modifications, like introducing notches[15] and steps[16], where the domain wall has higher probability to stop. However, even more advanced triangular notches[13] are unable to prevent overshoot for large current. This problem is aggravated by the high domain wall motion speed requirement for practical applications. To the best of our knowledge, there have not been any experimental reports of error-free domain wall memory with geometrically defined pinning regions. Other non-geometrical ways to implement the periodic energy pattern include chemical modification and ion implantation of the magnetic materials[17,18]. Although they may offer a larger tunable energy range, they require more complex fabrication techniques, and suffer from the same problem as the geometrical methods: the drive current pulses must be very precisely controlled. In a recent report[19] an error-free domain wall shift register was presented based on different principles. Here, regions with alternative net SOT signs were created, so that only when the applied current reverses, the domain wall progresses to the next section. The disadvantage of this design is that for multi-level memory applications, the top of the magnetic channel needs to be in contact with the magnetic tunneling barrier (MgO), but to change the net SOT sign, additional material must be deposited on top of the magnetic channel, which makes the multi-level memory operation impossible.

The paper is organized as follows: In Section 2, we will introduce the concept of a multi-bit domain wall motion magnetic memory based on a meandering magnetic channel. In Section 3, we introduce the details of the micromagnetic simulations, and in Section 4, we show a simple device design, followed by the possible material choices in Section 5. In Section 6 we will introduce the synthetic antiferromagnetic layer as a mitigation for the shortcomings in the domain wall motion control for the design discussed in Section 2. We will then introduce in Section 7 a new design that will allow bi-directional switching and improved scalability. The design and underling physics of the pinning layers used in the new design is discussed in detail in Section 8. We close with a discussion of the work presented and a brief conclusion in Sections 8 and 9.

## 2  Domain-wall motion multi-bit memory

We first propose a meandering magnetic channel with smoothly varying width dimension mitigating the drive current control problem. The basic operation relies on the fact that the force driving the domain wall is related to the current direction normal to the domain wall $F \sim \bm{j} \cdot \bm{n}$, taking a simple approximation. For simplicity we will assume, for now, that the domain wall will move perpendicular to the channel cross section (we will include a domain wall tilt later). For more accurate understanding, the dynamical equations of the domain wall can be written down with the help of Lagrangian mechanics with collective coordinates, and with some reasonable simplifications, and then verified by micromagnetic simulation[20–23]. In a simple strip channel, shown in Figure 2a, the domain wall will directly respond to the drive current and will therefore be exposed to all its uncertainty. The domain wall can move forward or backward depending on the direction of the current. Once the stimulating current pulse is complete, uncertainty in the domain wall position remains. A meandering channel with 180° turns, shown in Figure 2c and 2b, will force the domain wall to stop at the cusp of the turn because the driving force is zero. However, thermal fluctuations and material defects will provide fluctuations around this point that allow the domain wall to respond to either current direction.

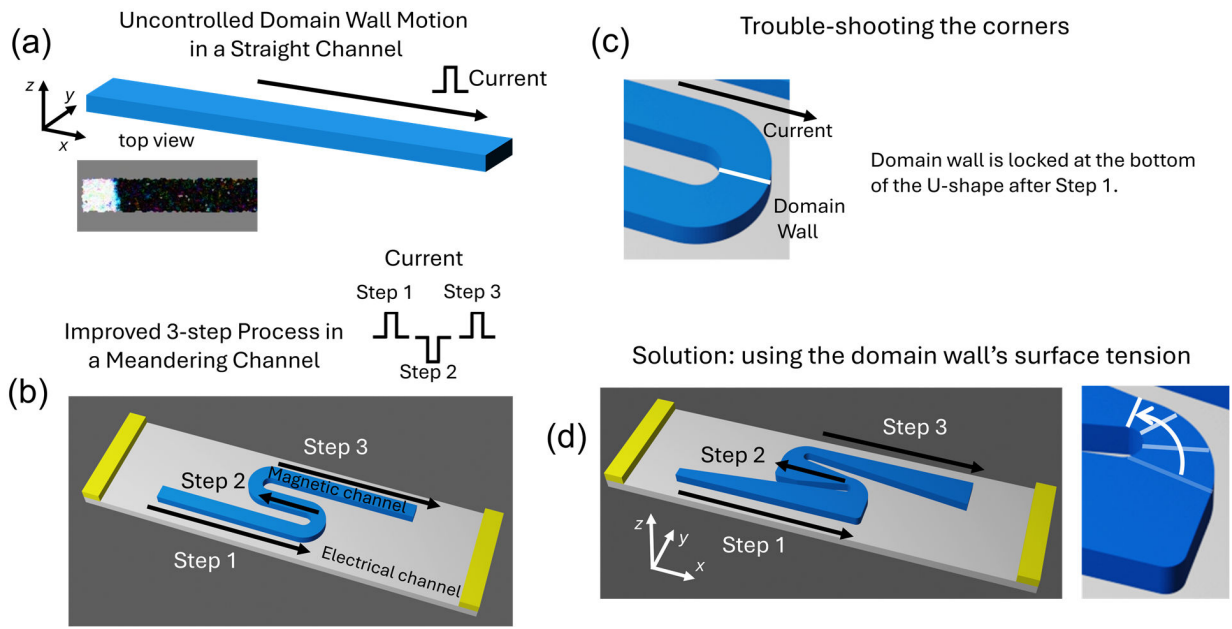

Figure 1. Fundamental ideas behind our meandering domain wall channel design. (a) A conceptual illustration of an uncontrolled domain wall motion in a straight channel. The inset is an example micromagnetic simulation of such a situation. The white region is where the magnetic moments point outwards, while the black region is where the magnetic moments point inwards. The effects of finite temperature and realistic materials and lithography defects have been considered. (b) A conceptual illustration of the improved 3-step switching process in a meandering channel. Current pulses are strong and long enough to move the domain wall

to the bottom of the U-turns. (c) Troubleshooting the corners. The domain wall cannot easily pass the U-shaped bottom because it does not have significant inertia to go to the other side. When a large current is applied, the stable position of the domain wall is the exact bottom of the U-turn. So, as the current reverses direction, the domain wall does not move to the other side. (d) By giving the U-turn parts of the channel an asymmetric shape, the domain wall's surface tension is strong enough to push it through the U-turn automatically, once the widest part of the U-turn channel has been overcome.

We now introduce a second design element, a taper in the magnetic channel downstream of the cusp that reduces channel cross section area going forward (Figure 1d). This offers three advantages. First, the surface tension of the domain wall pushes it forward[24] because of the reduction in the domain wall surface area, so that the equilibrium position is when the domain wall becomes parallel to the *x* axis. Second, because the force on the wall is now reversed, it localizes the domain at the bottom of the U-shape, regardless of current overdrive. Third, it places the domain wall in a position where current reversal can now move the domain wall to the next hairpin bend in the -*x* direction (Figure 1b). The exact shape of the neck region is further carefully optimized through micromagnetic simulations as discussed later.

With the modified corner the domain wall motion is controlled, the domain wall can pass the U-shaped neck, moves from the right to the left, and travels to the next U-shaped neck. As the current is reversed again, the domain wall passes the second U-shape neck and travels to the third, and so on. Due to the meandering structure, only by reversing the current, the domain wall is advanced, eliminating the overshooting problem. We can therefore turn the domain wall motion into an intrinsically digital, step-by-step motion, with the help of a geometrically latching design, which is fully operational with electric current, and no external field is needed, providing the base for a multi-bit magnetic memory.

Readout can be done by depositing an insulating layer of MgO and a fixed magnetic layer over the magnetic channel (see Figure 3), making the entire device a magnetic tunnel junction that generates a well-defined magnetoresistance[25]. With high quality PMA materials and MgO layer[26], at room temperature, the magnetoresistance can be as high as 100%.

## 3    Micromagnetic device simulations

Detailed micromagnetic simulations were carried out to detail out the basic device concept described above. A demonstration of a test device using the meandering channel principle. The simulation was performed on MuMax3, a GPU-accelerated open-source micromagnetic simulation package[27]. The materials parameters used are as follows. Cell size (x, y, z) = 2 nm by 2 nm by 0.6 nm. Saturation magnetization $M_S = 1.1 \times 10^6$ A/m, magneto-crystalline PMA constant $K_0 = 1.27 \times 10^6$ J/m$^3$, net PMA constant $K_{an} = K_0 - \frac{1}{2}\mu_0 M_S^2 = 5.097 \times 10^5$ J/m$^3$, net PMA field $B_{an} = \frac{2K_{an}}{M_S} = 0.9268$ T, ferromagnetic exchange stiffness $A_{ex} = 16$ pJ/m, interfacial DMI exchange constant $D_{ex} = 1$ mJ/m, Gilbert damping constant $\alpha = 0.02$. The magnetic material is modeled after CoFeB ($Co_{40}Fe_{40}B_{20}$), and the SOT source material is modeled after tungsten (metastable β-W, common for thin films). with an effective spin Hall angle of 0.30. When growing CoFeB on W, a dead layer without any magnetic moment will form, and the thickness is usually between 0.5-0.8 nm. The thickness of CoFeB (0.6 nm) and the saturation magnetization considered here are obtained after the dead layer was excluded. We note that the perpendicular anisotropy of CoFeB is heavily dependent on its thickness and its seeding and capping layers. In this case, we assume the stacking structure to be W/CoFeB/MgO, where the seeding layer is W and the capping layer is MgO, so $K_{an} = 5 \times 10^5$ J/m$^3$ is a reasonable value. Material parameters were summarized from the literature[26,28–33], on the optimistic side.

# 4 Practical device considerations

To initialize the domain wall, a region of defined magnetic orientation must be created at the beginning of the chain. This can be done in various ways[14]. One way is to use a local Oersted field (reference) to initiate a domain, which has been extensively studied theoretically[34], and another way is to use a local MTJ and STT to initiate a domain, and the third way is to begin the channel with a hard magnetic material as a fixed seeding domain. While the first two methods allow repeated reset at the beginning of the channel, the third method fixes the magnetic orientation permanently.

For device fabrication, the greatest challenge so far is to make the magnetic channel and material smooth enough to enable the graduated U-shaped neck. If the U-shaped necks work properly, the device should work deterministically and reliably. There are mainly two factors determining the pinning: shape defects created by the lithography patterning process, and material defects created by the materials deposition process[14]. It is crucial to verify that the materials and patterning quality requirements are achievable within the current technological limit. This can be done by numerical micromagnetic simulations with realistic temperature and defect distributions. Micromagnetic simulations suggest that the magnetic device needs to be very small and be patterned with very smooth edges, and low coercivity and low pinning field materials like CoFeB should be preferred[35–37]. However, given the correct conditions, the design has been shown to work reliably and there are no fundamental challenges that invalidate the design.

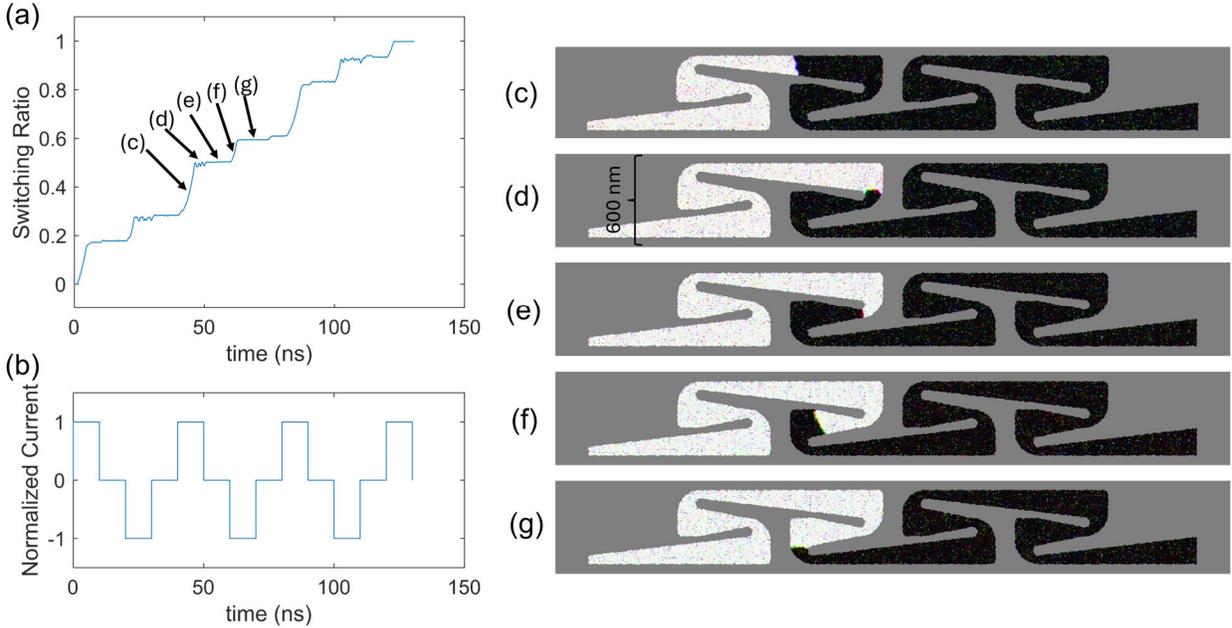

Figure 2. The domain wall is first nucleated at the left end of the device, and then propagates in the magnetic channel, driven by the spin current generated by the heavy metal SOT layer below the magnetic channel. (a) From micromagnetic simulation, we observed switching ratios from a subset of the meandering channel. (b) Normalized applied current in the electric SOT channel, which is underneath the magnetic channel. The current pulses have a density of $2.5 \times 10^7$ A/cm$^2$ in the tungsten layer. (c)-(g) are snapshots of the test device's magnetic states. The steps are non-uniform: larger steps correspond to longer segments of the channel, while smaller steps correspond to shorter segments. These are top view micromagnetic simulation results: (c) is during the long segment switching. (d) is when the current stays on, and the domain wall is locked at the bottom of the U-shape without the ability to further move to the right. (e) is when the current switches off, and the domain wall finishes the U-turn with the help of surface tension, and it is locked at the small neck, preparing for the next driving current. Similarly, (f) is when the domain wall is moving for the next step, and (g) is when the domain wall is locked at the next corner of the channel while the current is still on.

## 5 Material choices

The basic concept of our meandering channel principle is verified in the micromagnetic simulation sample [Figure 2] and the full MTJ stack is shown in Figure 3a and Figure 3b.

For efficient data readout, the best material for the tunneling barrier is MgO, and CoFeB is the industry standard for ferromagnetic free layers[14,38,39]. CoFeB is chosen for its high-quality perpendicular anisotropy, low damping constant, high tunnel magnetoresistance with MgO, good manufacturing compatibility (deposited by sputtering at relatively low temperatures), and tunable magnetic properties[14,38,39]. For smooth domain wall motion, it is critical to reduce the material defects so that the domain wall can travel freely with a small driving force. However, a disadvantage of CoFeB is its relatively low speed of domain wall motion, experimentally demonstrated[40,41] up to 15 m/s, compared to some other materials like Co and Co/Ni, which can go up to a few hundreds of m/s[14,42].

As for the underlayer (the current drive layer), it must satisfy three requirements[14,38,39]: (i), it gives CoFeB a high-quality perpendicular anisotropy; (ii), it has strong spin-orbit coupling, and generates large enough spin current, preferably from the spin Hall effect; (iii), it induces a strong enough interfacial Dzyaloshinskii-Moriya interaction in CoFeB, which is crucial to support the Néel-type magnetic domain wall in CoFeB and a faster domain wall motion velocity.

Table 1. Practical Choices of SOT Source Layer Material[28,29]

|  | Pt | β-Ta | Best Choice: β-W |
|---|---|---|---|
| Spin Hall Angle $\theta_{SH}$ | 0.07 | -0.12 | **-0.30** |
| Resistivity $\rho$ (μΩ·cm) at 5 nm thickness | 30 | 200 | **200** |
| Critical Current $J \propto \frac{1}{|\theta_{SH}|}$ ($10^8$A/cm$^2$) | 4.3 | 2.5 | **1.0** |
| Power Consumption $P \propto \rho J^2 \propto \frac{\rho}{\theta_{SH}^2}$ (Normalized with β-W) | 2.8 | 6.3 | **1.0** |
| Perpendicular Magnetic Anisotropy (PMA) as a seeding layer for CoFeB | Difficult to get high quality PMA | Easy to get high quality PMA | **Easy to get high quality PMA** |
| Dzyaloshinskii-Moriya Interaction (DMI) constant for PMA CoFeB (mJ/m$^2$) | 1.0 | 0.05 | **0.5** |
| Sufficiency of supporting fast-moving Néel-type domain wall | Yes | No | **Yes** |

Currently, three metals have been widely studied for their large spin Hall effects: Pt, β-Ta, and β-W. All of them are stable enough for the typical back-end-of-line process. We will give a discussion of these materials and decide the best material. The data were summarized in Table 1. Overall, it seems that β-W can be the best choice for our device, because it is a good seed layer for high-quality PMA CoFeB, and it allows the lowest device power consumption, and it gives strong enough DMI constant to support the Néel-type magnetic domain wall which can move fast.

While the basic design principles work, i.e., the domain wall moves as expected according to the channel U-turns and the switching ratio can be converted to a MTJ conductance state, we observe two shortcomings in operation and geometrical design. (1) the driving current sometimes causes irregular oscillations of the resistance state, especially when the domain wall is in a right-turning U-turn, and the irregular oscillations only stop when the driving current stops. (2) The conductance states are not spaced equally. In a multi-level memory, it is preferred that the electrical conductance states are equally spaced. But here, a certain conductance stage comes from the TMR value, and it is determined by the total areas of the "up" and "down" domains at that stage. As the domain wall travels along different segments of the magnetic channel, the segments with larger areas will have larger contribution to the conductance states compared to the segments with smaller areas, and therefore give unequally spaced conductance states.

Furthermore, the device architecture only allows one sided switching, which means the conductance can only be changed in one direction. To realize an incremental change in conductance, we need to move to a pair of channels and employ differential sensing, like that discussed in the application of PCM devices for neuromorphic computing[43]. The disadvantage, however, is that eventually, one of the channels will saturate, and reset of the pair is required. This can be accomplished through local MTJ at the beginning of the channel that is also used for the initialization, as discussed above. Although this process is feasible, it might negatively impact the performance, due to the reset operation.

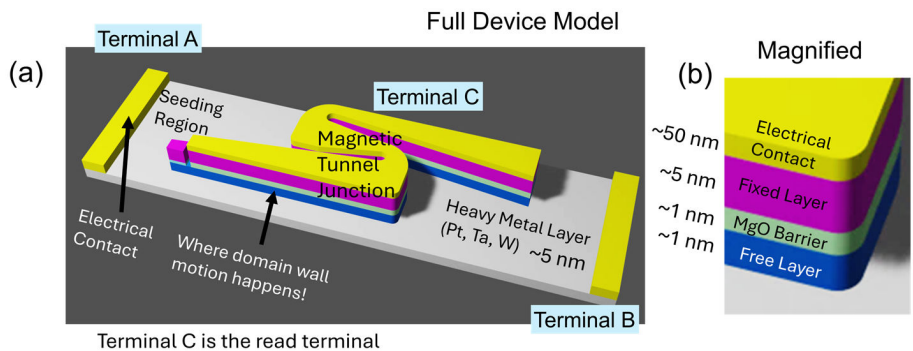

Figure 3. Practical considerations of such test devices. (a) A test device as introduced in Fig. 2 needs to have a seeding region for the domain wall to nucleate, and a magnetic tunnel junction should be fabricated over the main region of the channel. (b) A closer view of the magnetic tunnel junction stack.

## 6  Synthetic antiferromagnetic layer

To improve the switching characteristic further, we introduce synthetic antiferromagnets to replace ferromagnets of the MTJ free layer as shown in Figure 4. Synthetic antiferromagnetic (SAF) layers are made of ferromagnetic layers antiferromagnetically coupled with each other, with the help of spacer metals such as Ru [44]. We will now discuss the critical value of SAF layers to our device design. Due to the antiferromagnetic coupling between the magnetic layers, the strong exchange coupling torque dramatically changes the magnetic dynamics of the magnetic layer resulting in a faster switching rate which affects the domain wall motion as well[23,44–46]. In addition, compensated synthetic antiferromagnets mitigate domain wall shape tilting which will lead to a more reliable motion control of the domain wall[23,46].

Assuming no defects exist in the system, and the temperature is absolute zero, without applying any external magnetic field or current, the domain wall shape should be a perfectly straight line when looking at the nano-strip from the top, as the straight line is perfectly perpendicular to the nano-strip edges. However, once the domain wall starts to move, as shown in Figure 5, driven by either an out-of-plane magnetic field or an electric current along the nano-strip, the domain wall shape experiences a tilt. This effect is widely known in theoretical[23] and experimental studies[47]. The larger the driving force is, the more tilted the domain wall becomes[23]. Driven by the current, the "up-down" [Figure 5(a)], and "down-up", [Figure 5(b)], domain walls move in the same direction, but tilt in the opposite ways. This tilt may interfere with the device function, so it is often undesirable.

When changing the ferromagnet to a compensated synthetic antiferromagnet, by simply putting a spacer layer of strong Ruderman-Kittel-Kasuya-Yosida (RKKY) interaction, and another ferromagnetic layer identical to the original one, as the bottom domain wall is "up-down", the top domain wall becomes "down-up". As we now have "up-down" and "down-up" domain walls in the identical bottom and top layers respectively, and they move in the same direction, domain wall shapes want to tilt in opposite directions, but the domain walls are coupled, so they also want to tilt in the same direction. The final result is that the domain wall remains nearly perpendicular to the side walls[23]. The mitigation of domain wall tilting by compensated synthetic antiferromagnets has been reported theoretically[23] and experimentally[46], and it also has been confirmed by our micromagnetic simulations in various conditions.

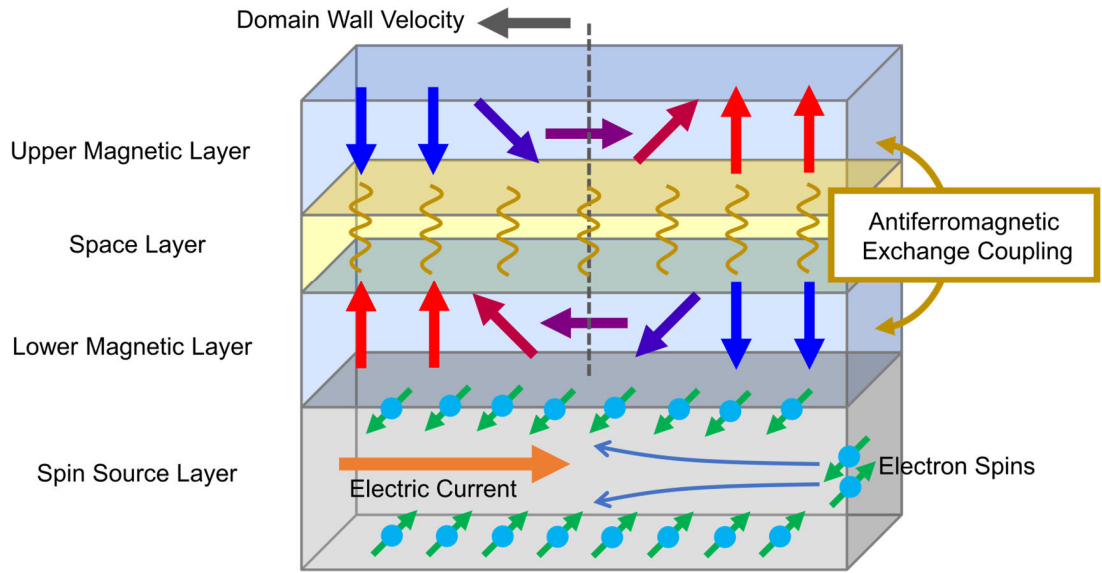

Figure 4. Schematic of SOT-induced domain wall motion in a synthetic antiferromagnet.

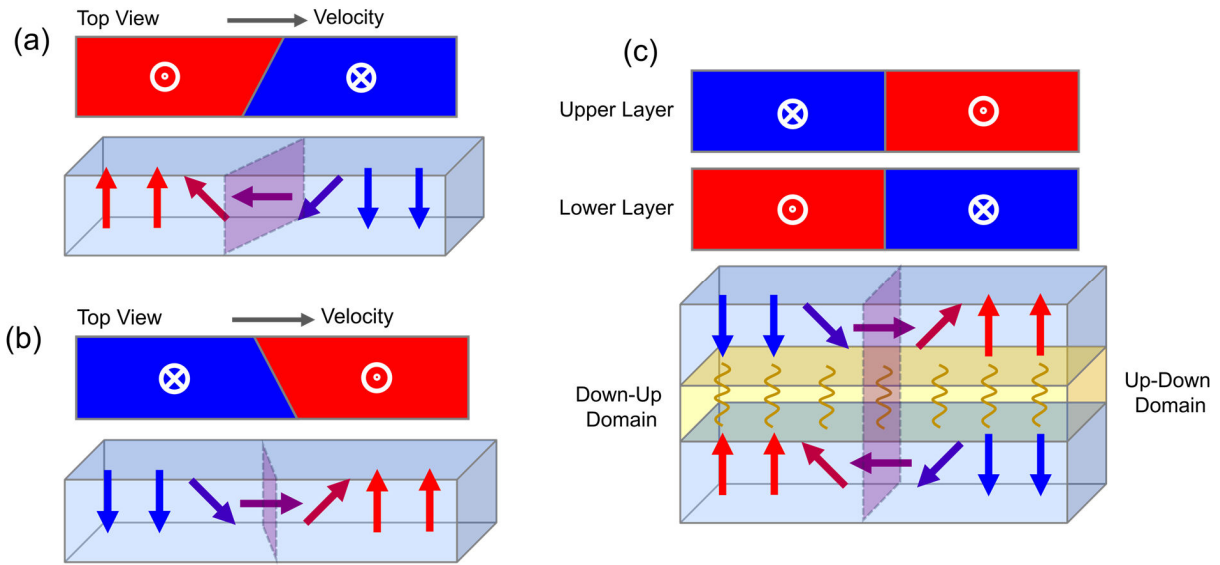

Figure 5. The tilting in the width direction of the channel for PMA ferromagnetic or antiferromagnetic domain walls. (a) and (b), for ferromagnetic domain walls, depending on the domain wall polarity and velocity direction, there are two possible tilting directions, which have been repeatedly observed in our previous micromagnetic simulations. (c) For synthetic antiferromagnetic domain walls, if the upper and lower layers are decoupled, they have opposite tilting directions, but as they are coupled together, their tilting directions largely cancel out, showing negligible overall tilting. This is beneficial to the precise control of our devices.

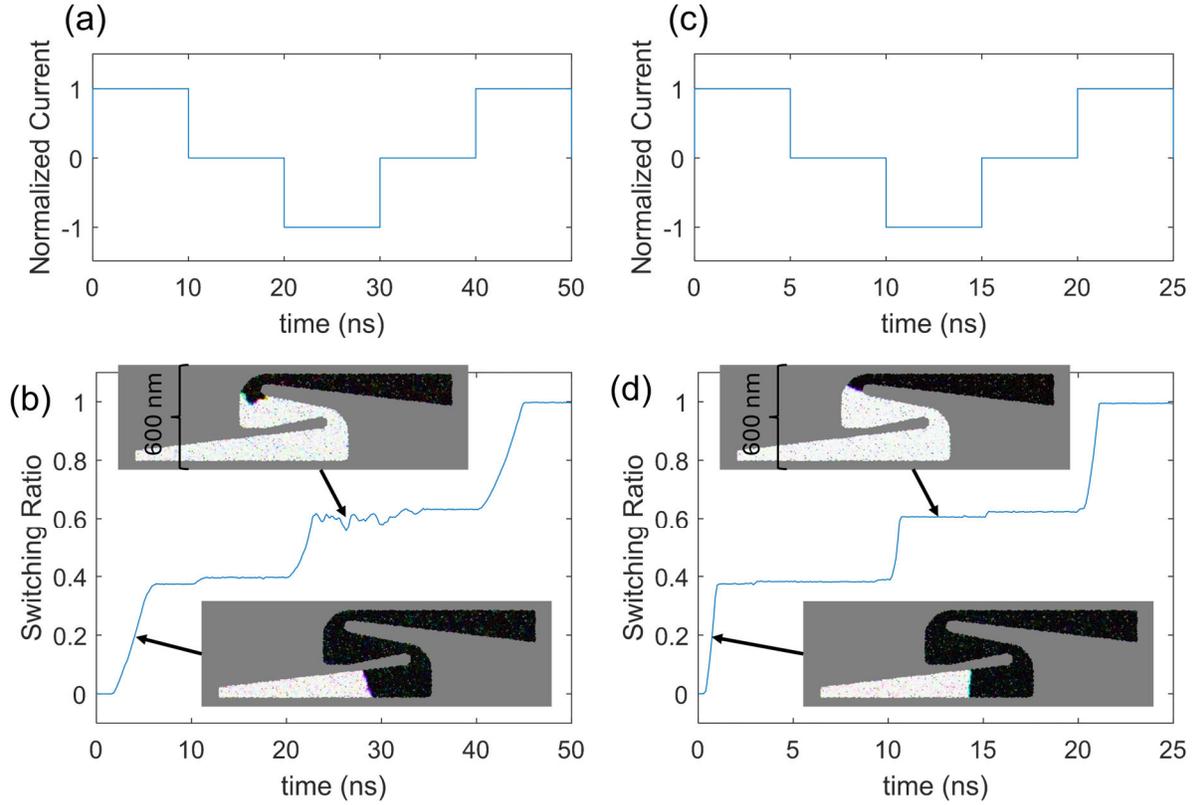

Figure 6. The comparison between simple test devices of identical lithographical geometry, made of ferromagnetic and synthetic antiferromagnetic layers, respectively. (a) The driving current for the ferromagnetic device. (b) The switching plot for the ferromagnetic device. (c) The driving current for the synthetic antiferromagnetic device. (d) The switching plot for the ferromagnetic device. The switching becomes several times faster, and cleaner, with much lower domain wall tilt and random oscillations.

The replacement of the ferromagnetic free layer by a SAF stack significantly improves the switching speed and accuracy of the device, as we show for the two-way switching device in Figure 6. The speed increase is well-known, attributed to the large interlayer exchange interaction stabilizing the domain wall and extending its steady motion regime to higher speeds[23]. In this case, the RKKY interlayer exchange coupling constant $J_{\text{RKKY}} = -1.6667 \times 10^{-3}$ J/m$^2$, a large but feasible value[48], which corresponds to an antiferromagnetic exchange coupling field of $\mu_0 H_{\text{ex}} = -\frac{J_{\text{RKKY}}}{M_s t} = 2.53$ T. As corner turning is still steered by surface tension of the domain wall, the device is relatively sensitive to defects. Defect size and density must be smaller than a certain threshold for the device to work. Because surface tension is only effective at a smaller width dimension, so that devices need a channel width of ~ 100 nm or less. And one additional advantage is that the thermal stability becomes doubled, due to the doubled total magnetic layer thickness, which leads to better device miniaturization.

# 7 Bidirectionally switching device design

The introduction of SAF makes the domain wall motion control more reliable. We will now tackle the problem of bidirectional switching. Bidirectional switching is important for multi-bit applications because a new memory state can be set by modifying an existing one without prior erase operation. To this end we propose an alternative design shown in Figure 7. Instead of winding the magnetic domain wall channel into a strip with two preferred directions, we wind it into a 2-dimensional pattern with four preferred directions. The result is a four terminal cell where current can flow in four different directions. This gives us a much greater flexibility of moving the domain walls.

Although the basic principles remain the same, because the domain wall only turns 90° instead of 180°, corner turning is steered entirely by applied current across the four terminals, instead of the surface tension of the domain wall. This important change makes the device relatively insensitive to defects, and the device works at small dimensions and large dimensions. It is also easy to steer the domain wall backwards by reversing the applied current directions. Furthermore, the device can be smaller and the switching steps shorter, and the shape requires less precision, and conductance states are evenly distributed.

Since we have well-defined start and end points of our meandering channel, we can achieve magnetic pinning by antiferromagnets, or simply by high-coercivity ferromagnets, to initiate a domain wall at the start, and we can block the domain wall at the end to prevent it from disappearing. Another advantage is that the shape of the meandering channel can become much simpler, and the switching steps significantly shorter.

However, there are also disadvantages of this new design: the heavy metal channel width increases, and the required current may be higher to maintain the required current density. Changing the conductance state of the cell could be done by a read before write, to determine the correct signals to write to the desired new bit state.

In the following example, we designed an error-free meandering domain wall memory with 16 conductance states as shown in Figure 7. Each switching step has the same shape and length which means the conductance states are equally spaced. The resistance states are named from 0 to 15 sequentially.

While in the earlier implementation surface tension was used to help the domain wall around the corner to respond to the reversed current direction, this modified design eliminates the need for surface tension but still requires the domain wall to stop accurately when the channel makes a 90° turn. If we use a single ferromagnetic layer for the domain wall motion, the domain wall has a pre-existing tilt, as discussed above, when traveling in the straight channel along the current, when the tilting direction is opposite to the corner turning direction, the domain wall needs to turn for more than 90 degrees, and it is more difficult. This usually triggers turbulent motion of the domain wall around the corners and results in unpredictable behaviors, that the domain wall will not stop at the designated location. However, with the introduction of synthetic antiferromagnets, this problem is eliminated. Suppose we have a well-designed, compensated, strongly coupled synthetic antiferromagnet. As the domain wall moves with negligible tilt, the domain wall can safely pass all corners precisely, according to the driving current's command because at any 90-degree turn the driving force will be zero and the domain wall motion will stop. This has been confirmed by our micromagnetic simulations under realistic conditions [Figure 8].

Different from the simplified geometry in Figure 7, to precisely stop the domain walls at the corners of the device, additional pinning sites must be engineered to mitigate the influence of thermal fluctuations. The core idea is to allow the domain wall to move forward. We will discuss the design criteria for the pinning sites below. The improved switching characteristics are shown in Figure 8.

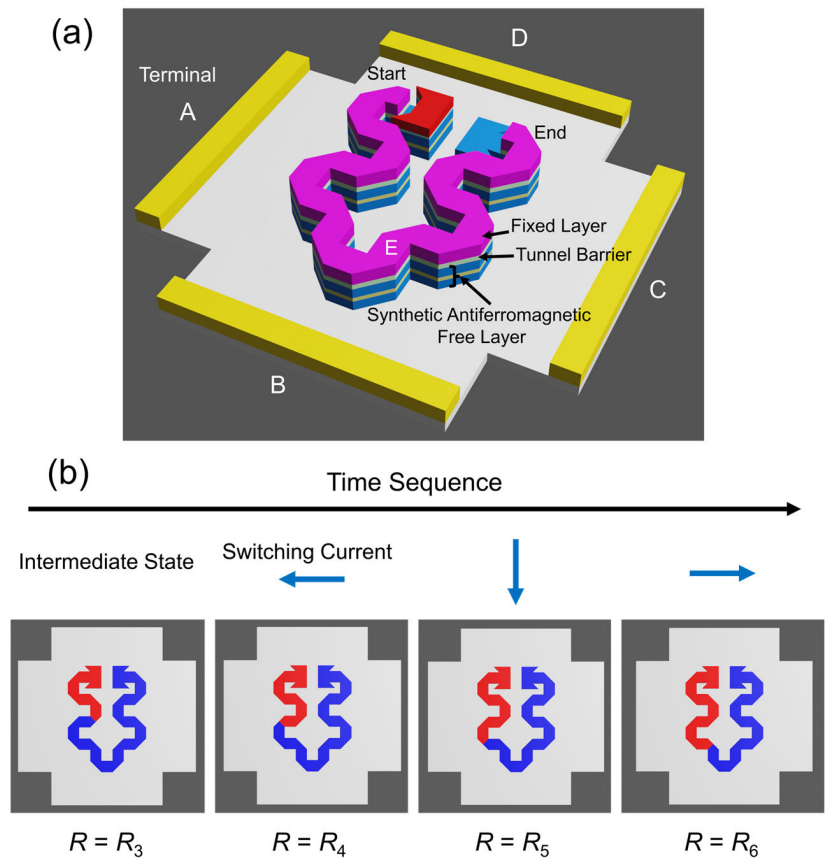

Figure 7: Conceptual schematic of 4-way switching device. (a) 3D illustration of the concept. (b) Working principle of the device, looking from the top. Only the top layer of the synthetic antiferromagnetic layers is shown. Red means magnetization pointing outwards, and blue means magnetization pointing inwards.

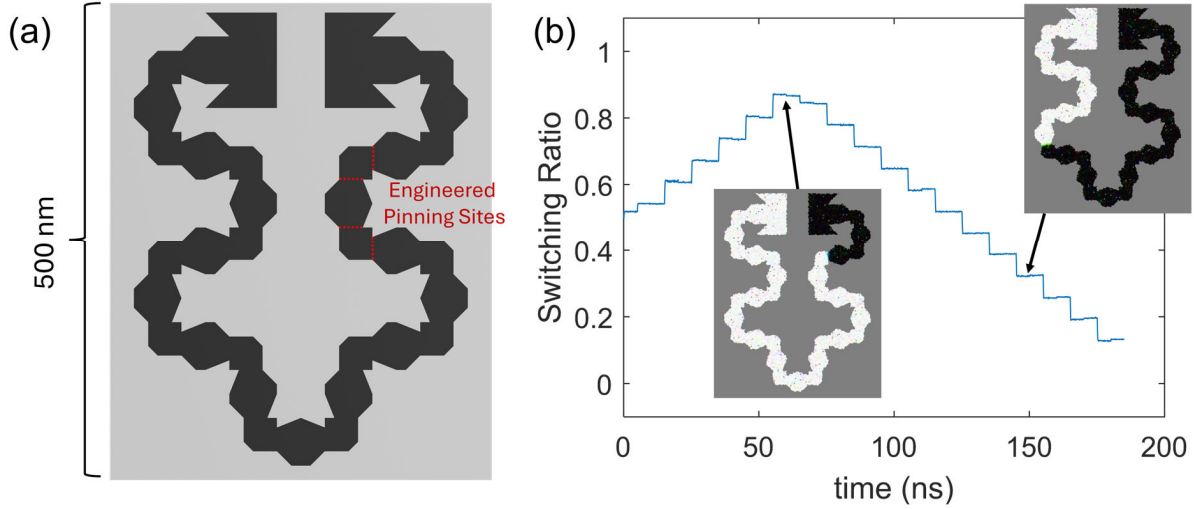

Figure 8: An implementation of 4-way switching device. The SAF magnetic free layer is made of two PMA CoFeB layers with an effective thickness of 0.6 nm for each layer, and a Ru space layer with a thickness around 0.7 nm. The single CoFeB layer is assumed to be identical to what we have seen in the previous simulation with a single ferromagnetic free layer. The SOT source layer is β-W, with an effective spin Hall angle of 0.30. The switching current is $8.33 \times 10^7$ A/cm². (a) The geometric shape of the 4-way switching device's magnetic channel with engineered pinning sites. (b) Testing of the 4-way switching device using micromagnetic simulation. The inset figures are the magnetic states of the upper layer during the operation. By controlling applied current directions, the device resistance states can be tuned freely and reversibly.

Table 2

|  | Test Device | Improved Version |
| --- | --- | --- |
| **Number of states** | 16 | 16 |
| SAF Free layer thickness t | 1 nm/0.7 nm/1 nm | 2 nm/0.7 nm/2 nm |
| Perpendicular anisotropy $K_{eff}$ | $5 \times 10^5$ J/m³ | $1.25 \times 10^6$ J/m³ |
| Spin-orbit torque efficiency $\xi$ | 0.3 (β-W) | 1.0 (emergent materials) |
| Materials and nanofabrication quality | Normal | High |
| Largest/smallest channel width w | 50 nm/25 nm | 10 nm/5 nm |
| **Device center size** | **500 nm by 500 nm** | **100 nm by 100 nm** |
| Practical write current density (roughly two times the critical value) | $1.5 \times 10^8$ A/cm² | $1.5 \times 10^8$ A/cm² |
| Total current | 3750 μA | 750 μA |
| **Write time (change for one step)** | **1 ns** | **0.2 ns** |
| Write power (device center) | 5.6 pJ | 45 fJ |

There are three key components of device performance: size, speed, and power consumption. We will discover the limiting factors in this device design, and then summarize our reasonable short-term and long-term goals of device performance.

# 8 Pinning layer design

The size of the device is fundamentally determined by the dimensions of the domain wall: width (equal to the channel width), height (equal to channel film thickness), and thickness. Since its height is only a few nanometers, we can assume that the magnetic moment distribution along the height is uniform within any ferromagnetic layers and antiparallel between the antiferromagnetically coupled ferromagnetic layers. The domain wall thickness is the horizontal distance where the magnetic moments gradually reverse directions. It is not defined by the device geometry, but by the balance among exchange, anisotropy, and demagnetizing energies.

For simple ferromagnetic layers, both Bloch walls and Néel walls have a characteristic width $d = \pi\delta = \pi\sqrt{\frac{A_{ex}}{K_{an}}}$, where $A_{ex}$ is the exchange energy constant and $K_{an}$ is the anisotropy energy constant[20]. For $A_{ex} = 20$ pJ/m and $K_{an} = 5 \times 10^5$ J/m$^3$, $d = 19.9$ nm. This poses a lower bound on the lateral dimension of the magnetic device channel. Secondly, the domain wall should have enough thermal stability when it is pinned to a certain location. Because we introduce pinning potentials by variable channel width, the criterion should be that the energy difference between the bottom and the edge of the potential well is greater than a certain thermal stability factor, so that the chance that the domain wall escapes the potential well is extremely small, at the upper limit of the operating temperature and throughout the designed life span. In our case, this energy difference is defined as the energy difference between the domain wall at its minimum length, and the extended domain wall at the verge of escaping the pinning.

To understand the domain wall energy, we must first investigate what makes a Néel wall, and how we can stabilize it. This is because only a Néel wall can be efficiently driven by SOT in a magnetic ultrathin film with PMA. In most cases, it is reasonable to assume that the domain wall width $d$ is much larger than the film thickness $t$, and the magnetic channel width $w$ is larger than $d$. The energy difference between a Bloch wall and a Néel wall can be regarded as the competition between the demagnetizing energy and the DMI energy, because the ferromagnetic exchange energy and the anisotropy energy are the same between Bloch and Néel walls. In the case of $w \gg d$, the Bloch wall's total energy can be written as $E_{\text{Bloch}} = tw\left(4\sqrt{A_{\text{ex}}K_{\text{an}}}\right)$, without any additional energy contributions, while the Néel wall's total energy is approximately $E_{\text{Néel}} = tw\left(4\sqrt{A_{\text{ex}}K_{\text{an}}} - \pi D_{\text{ex}} + \frac{\ln 2}{\pi}\mu_0 M_S^2 t\right)$, where the added two terms come from DMI and demagnetizing energy, respectively[20,49–51], and $D_{\text{ex}}$ is the DMI exchange constant, and $M_S$ is the saturation magnetization of the ferromagnetic layer. When $E_{\text{Néel}} < E_{\text{Bloch}}$, or $\pi D_{\text{ex}} > \frac{\ln 2}{\pi}\mu_0 M_S^2 t$, the Néel wall can be stabilized, which requires a large enough $D_{\text{ex}}$ and a small enough $t$. $D_{\text{ex}}$ can go up to 3 mJ/m² in some materials systems. If $D_{\text{ex}}$ is as low as 0.25 mJ/m², the maximum $t$ is around 2.4 nm. And when $w$ reduces, $E_{\text{Bloch}}$ will further rise due to the demagnetizing energy from the width sides of the domain wall channel, but $E_{\text{Néel}}$ will decrease slightly, making the Néel wall more favorable. And with considering synthetic antiferromagnetic free layers, the energy of Néel wall will be further lowered compared to the Bloch wall, because the magnetic poles across the domain walls of the upper layer and the lower layer are the opposite, and thus reduce the magnetic dipolar energy.

Now we are ready to estimate the condition at which the domain wall can be reliably pinned at a certain location in the magnetic channel: the energy difference between the bottom and the edge of the potential well should be greater than a certain thermal stability factor. For the estimations here, we considered a modest improvement of our simulated W/CoFeB/MgO device made of a single ferromagnetic layer. $t = 1$ nm (not counting the dead layer, so it is slightly thicker than our simulation), $A_{\text{ex}} = 20$ pJ/m (slightly stronger than our simulation), $D_{\text{ex}} = 0.5$ mJ/m² (weaker than our simulation), $K_{\text{an}} = 5 \times 10^5$ J/m³ (the same as our simulation), $4\sqrt{A_{\text{ex}}K_{\text{an}}} = 1.2650 \times 10^{-2}$ J/m², while $\pi D_{\text{ex}} = 1.5708 \times 10^{-3}$ J/m². The best case would be to get $\pi D_{\text{ex}} - \frac{\ln 2}{\pi}\mu_0 M_S^2 t$ to be slightly larger than zero. Therefore, we can use $E_{\text{DW}} = 4tw\sqrt{A_{\text{ex}}K_{\text{an}}}$ as an approximation for the domain wall energy.

We set the required thermal barrier to $60k_B T$, a reasonable value according to industrial standards[38], which means the total error rate is around $3 \times 10^{-10}$, if the device is kept at 85°C for 10 years, assuming the switching attempt frequency $f_0 = 1$ GHz. This is determined by the Arrhenius Law: $f = f_0 e^{-\frac{\Delta E}{k_B T}}$. Assuming the only way to contribute to this thermal barrier is the difference of channel width, for W/CoFeB, $\Delta E_{\text{DW}} = 60k_B T \rightarrow \Delta w = 23.5$ nm. Given the geometrical limitations, we assume that $\Delta w$ takes up to half of the full width $w$, which gives $w \sim 50$ nm. But this value is only for a ferromagnetic free layer. If a synthetic antiferromagnetic free layer is made of two CoFeB layers, each of which has an effective thickness of 1 nm, and they are coupled by a Ru layer, the total thermal stability factor will double, which further shrinks the minimal full channel width ($w$) down to 25 nm. However, the limiting factor would then become the domain wall width $d$, and possibly the lithography patterning. So, in this design, we keep $w$ to 50 nm, so the device has twice as much as the required thermal stability factor. And as for switching current density, considering the updated thickness values, and scaling from our simulations, a current density of $1.5 \times 10^8$ A/cm$^2$ is more than enough to reliably move the domain wall. And as for the device area, when counting the center area for 4-directional current flow, it is about 10 times the main magnetic channel width, which is 500 nm. Considering the electrodes around the center area will add to the total size of the device. And we assume that the synthetic antiferromagnetic domain wall travels at a speed of a few hundred meters per second, given optimized material quality. To reliably travel 100 nm, counting the acceleration phase and other possible delays, takes about one nanosecond. This suggests that the device has a write power consumption on the order of a few pico-Joule, for an increment or a decrement. One unique challenge of this device design is the large current density required, which is still an order of magnitude larger than what is currently conveniently available from deeply scaled CMOS technology.

For our future goal, to shrink down the device further, the domain wall width $d$ becomes an obvious bottleneck. Previously, the effective perpendicular magnetic anisotropy is assumed to be $5 \times 10^5$ J/m$^3$, which is already on the larger end of various reports of CoFeB with PMA[26,30–33]. Given that $E_{DW} = 4tw\sqrt{A_{ex}K_{an}}$ and $d = \pi\sqrt{\frac{A_{ex}}{K_{an}}}$, and that $A_{ex}$ is difficult to tune, the only way to shrink the domain wall size is to use a much larger $K_{an}$. This can be achieved by using L1$_0$ phase FePt and CoPt, Co/Pt and Co/Pd multilayers, CoTb and TbFeCo ferrimagnets, Co$_2$FeAl and Co$_2$MnSi Heusler alloys[14,38,39]. However, a small enough Gilbert damping is also required to keep the domain wall velocity high. They also need to form synthetic antiferromagnetic coupling with a similar adjacent layer. Given that CoFeB is preferred for a large TMR, the upper layer of the synthetic antiferromagnet should be a composite with a CoFeB finish layer ferromagnetically coupled to the original material with very high PMA. It is possible to increase $K_{an}$ by one order of magnitude to $5 \times 10^6$ J/m$^3$, so that $d$ can be reduced to 6 nm. However, we do not need to go to that extreme. Setting $K_{an} = 1.25 \times 10^6$ J/m$^3$ and increasing the individual ferromagnetic layer thickness from 1 nm to 2 nm should be enough to reduce $d$ to 6 nm. If the magnetic channel width $w$ does not change, $E_{DW}$ is multiplied by a factor of 3, so that a down-scaling of the magnetic channel width $w$ and the width difference $\Delta w$ is allowed, while keeping $\Delta E_{DW} = 60k_BT$. When pushing to the limit, $w = 10$ nm and $\Delta w = 5$ nm. This shrinks the device center square down to 100 nm by 100 nm. A domain wall in this system is packed with energy. To move it, the SOT must be stronger. By using a material with a SOT efficiency of 1.0 and a similar electrical resistance as β-W, possibly using topological materials or other exotic materials[7], and when the material's pinning defects are carefully tuned and reduced, and when the device geometry defined by lithography is highly accurate, it should be possible to keep the current density the same, at $1.5 \times 10^8$ A/cm$^2$. And due to much shorter magnetic channel segments, the single step write time can be reduced by a factor of five, down to 0.2 ns, and the energy consumption is on the order of a few tens of femto-Joules. Although the full device and the peripheral circuitry may require a significantly larger total energy consumption, the characteristics of our design have the potential to satisfy the pressing needs of non-volatile memory and neuromorphic computing.

# 9  Discussion

We have introduced a SOT-MRAM cell that is able to store multi-bit information by controlling the domain wall motion. The key for this control is geometrically latching its position at predetermined pinning regions. The two mechanisms we introduced in this work provide different degrees of fidelity at the cost of more complex drive current control but with increased flexibility. The micromagnetic simulations indicate feasibility of this concept. There are, however, a few issues that are still waiting for a solution before a practical implementation of this concept is possible. In this section we highlight the outstanding issues that warrant additional research and the possible benefit to be gained.

The performance estimates in Table 2 suggest a write time of 200 ps. It is slower than fast SRAM but faster than embedded DRAM which makes SOT-MRAM a suitable candidate for higher level cache provided the bit-density is superior to alternative solutions[10]. While SRAM, STT-MRAM and embedded DRAM can only store one bit per cell, SOT-MRAM is able to store multiple bits. Therefore, the cell size is amortized across several bits. According to the ITRS roadmap[52] the SRAM cell size will converge to about 0.01 μm$^2$ in the next decade due to physical scaling limits. For the scaled version of the SOT-MRAM we predict a cell size of 0.02 μm$^2$ for 4-bit capacity, assuming that the cell area is twice as large as the device center, which gives a cell-size per bit of 0.005 μm$^2$ which is a significant advantage in density. It is of the same order as the bit-cell size of an STT-MRAM, however, avoiding the complicated trade-off between endurance and write speed. Whether this advantage can be transferred to the array level is currently an open question because it depends on the ultimate array architecture and the detailed circuit solutions for write and read access. With respect to embedded DRAM, which requires a large storage capacitor to be integrated with the logic process, SOT-MRAM provides a cost benefit due to a less complex process and an additional bit-cell density advantage.

The low value of TMR requires a careful analysis of signal to noise margin to make sure the 16 states of the 4-bit memory can be securely distinguished. In contrast to the STT-MRAM where write and read current pass through the MTJ in SOT-MRAM this is not the case. Therefore, the MTJ conductance can be modified without impacting the write performance. A thicker tunnel barrier would decrease the base conductance and at the same time could reduce process variations for a better signal to noise margin. This reduction in conductance is only limited by maintaining a sufficiently high read current. The target conductance is determined by the array architecture and the application. For instance, for a pure multi-bit memory the signal to noise margin must be significantly larger than that for deep learning crosspoint array.

A more serious problem for the SOT-MRAM is the high write current. Moving the domain wall requires a critical current density. This critical current density is determined by the spin-orbit torque efficiency $\xi$ and the resistivity ratio of heavy metal to first ferro-magnetic layer to avoid current shunting. To securely move the domain wall, we assume a current density that is twice the critical current density. The parameters shown in Table 2 are based on our micromagnetic simulation. We find approximately the same critical current density for the test and the scaled case due to an increased domain wall energy in the later. Only further increase in spin-orbit torque efficiency $\xi$ would reduce the critical current density. Using a heavy metal layer of 5 nm thickness from our simulation we infer a write current of 3750 µA and 750 µA for the test case and the scaled case, respectively. Device technology will deliver a current of approximately 800 µA/µm within the appropriate cell dimension[52] would give an upper bound of < 400 µA and < 80 µA for the two configurations, way below the required values. This problem is not unique to the device proposed here but is a general problem for density scaled SOT devices[53]. The straightforward solution is to find materials with a significant higher spin-orbit torque efficiency $\xi$ than that used for the scaled memory cell. Topological insulators and semimetals show larger spin-orbit torque efficiencies, however, there is large scatter in the available data and more research is required[7]. In addition to the large $\xi$ a low enough resistivity is desired to avoid shunting which would rise the required current. The high write current will also require a localized current source on cell level to avoid unnecessary voltage drops and related energy losses in the metal lines of the array. A local cell level current source for two-fold or four-fold current steering would also considerably simplify the array architecture. A possible solution would be to build the SOT-MRAM cell on top of the control circuit either monolithically or as 3D stacked chip. Both material and architectural solutions need to be explored for a practical implementation of this cell concept.

# 10 Conclusion

We have introduced a SOT-MRAM memory with multi-bit capability by controlling the domain wall motion with interspersed pinning regions at predetermined positions in a magnetic channel. This motion is current-driven and otherwise field-free and the pinning regions are geometrical features in the magnetic channel. Our simulations have shown that precise control of the domain wall is feasible under realistic conditions including thermal fluctuation and material defects. We find that a synthetic antiferromagnetic layer improves the switching behaviors of the SOT multi-bit memory. We also discussed the remaining challenges for a practical implementation of this new concept of which reduction of the drive current for the domain wall is the foremost important component.

# Acknowledgement

This material is based upon work supported by the U.S. Department of Energy, Office of Science, for support of microelectronics research, under contract number DE-AC02-06CH11357.

The submitted manuscript has been created by UChicago Argonne, LLC, Operator of Argonne National Laboratory ("Argonne"). Argonne, a U.S. Department of Energy Office of Science laboratory, is operated under Contract No. DE-AC02-06CH11357. The U.S. Government retains for itself, and others acting on its behalf, a paid-up nonexclusive, irrevocable worldwide license in said article to reproduce, prepare derivative works, distribute copies to the public, and perform publicly and display publicly, by or on behalf of the Government. The Department of Energy will provide public access to these results of federally sponsored research in accordance with the DOE Public Access Plan. http://energy.gov/downloads/doe-public-access-plan